\documentclass[twocolumn,showpacs,preprintnumbers,amsmath,amssymb]{revtex4}

\usepackage{amsmath,amssymb,graphicx}
\newcommand{\bea}{\begin{eqnarray}}
\newcommand{\eea}{\end{eqnarray}}

\begin{document}
%%%%%%%%%%%%%%%%%%%%%%%%%%%%%%%%%%%%%%%%%%%%%%%%%%%%%%%%%%%%%%%

%%%%%%%%%%%%%%%%%%%%%%%%%%%%%%%%%%%%%%%%%%%%%%%%%%%%%%%%%%%%%%%
\title{Axion as a Cold Dark Matter candidate}%: Linear stage}
\author{Jai-chan Hwang${}^{1}$ and Hyerim Noh${}^{2}$}
\affiliation{
         ${}^{1}$Department of Astronomy and Atmospheric Sciences,
                 Kyungpook National University, Taegu, Korea \\
         ${}^{2}$Korea Astronomy and Space Science Institute,
                 Daejon, Korea \\
         E-mails: ${}^{1}$jchan@knu.ac.kr,
                  ${}^{2}$hr@kasi.re.kr}

%%%%%%%%%%%%%%%%%%%%%%%%%%%%%%%%%%%%%%%%%%%%%%%%%%%%%%%%%%%%%%%
\begin{abstract}

Here we generally {\it prove} that the axion as a coherently
oscillating scalar field acts as a cold dark matter in nearly all
cosmologically relevant scales. The proof is made in the linear
perturbation order. Compared with our previous proof based on
solutions, here we compare the equations in the axion with the ones
in the cold dark matter, thus expanding the valid range of the
proof. Deviation from purely pressureless medium appears in very
small scale where axion {\it reveals} a peculiar equation of state.
Our analysis is made in the presence of the cosmological constant,
and our conclusions are valid in the presence of other fluid and
field components.

\end{abstract}

%%%%%%%%%%%%%%%%%%%%%%%%%%%%%%%%%%%%%%%%%%%%%%%%%%%%%%%%%%%%%%%
\noindent 


%%%%%%%%%%%%%%%%%%%%%%%%%%%%%%%%%%%%%%%%%%%%%%%%%%%%%%%%%%%%%%%
\maketitle

\vskip .0cm
%%%%%%%%%%%%%%%%%%%%%%%%%%%%%%%%%%%%%%%%%%%%%%%%%%%%%%%%%%%%%%%
%
% Introduction
%
%%%%%%%%%%%%%%%%%%%%%%%%%%%%%%%%%%%%%%%%%%%%%%%%%%%%%%%%%%%%%%%
%\section{Introduction}

Cold dark matter (CDM), despite its unknown nature, became an
essential ingredient in the current cosmological studies concerning
the large-scale structure formation. CDM in a cosmological constant
dominated world model (often termed the $\Lambda$CDM model) is
currently the most successful candidate of cosmological models. The
nature of CDM and the nature of cosmological constant (or some other
dynamical dark energy), however, still remain as fundamental
mysteries of present day physical cosmology. From early days of dark
matter studies axion as a coherently oscillating scalar field is
widely accepted as a candidate for the CDM \cite{Axion-CDM}.
Confirmation of the case using the relativistic linear perturbation
analysis was made previously \cite{Ratra-1991,Axion-1997}, for a
recent study see \cite{Sikivie-Yang-2009}. In this work we present a
more general proof that the axion can be regarded as the CDM to the
linear order perturbation. We also derive an effective equation of
state of the axion which could be important in the solar system
scale if the system is in the linear regime. We set $c \equiv 1
\equiv \hbar$.

%%%%%%%%%%%%%%%%%%%%%%%%%%%%%%%%%%%%%%%%%%%%%%%%%%%%%%%%%%%%%%%
We consider the axion as a minimally coupled scalar field with $V =
{1\over 2} m^2 \phi^2$. The relevant current scales we are concerned
correspond to \bea
   & & {H \over m} = 2.133 \times 10^{-28} h \left( { m \over
       10^{-5} eV} \right)^{-1},
\eea where we set currently $H \equiv 100h km/(sec Mpc)$. In the
following analyses we strictly ignore ${H / m}$ higher order terms.
But we do not impose any limit on the wavenumber $k$; this is more
general than our previous work in \cite{Axion-1997} and
\cite{Sikivie-Yang-2009}. We {\it take} a spatially flat background
with the cosmological constant $\Lambda$ and the axion; inclusion of
$\Lambda$ is also more general than previous studies.

We consider the temporal average of the oscillating scalar field
contributes to the background fluid quantities, thus \bea
   & & H^2 = {8 \pi G \over 3} \mu
       + {\Lambda \over 3}, \quad
       \dot H = - 4 \pi G \left( \mu + p \right),
   \nonumber \\
   & & \mu = {1 \over 2} \langle \dot \phi^2 + m^2 \phi^2 \rangle, \quad
       p = {1 \over 2} \langle \dot \phi^2 - m^2 \phi^2 \rangle,
   \label{BG-fluid} \\
   & &
       \ddot \phi + 3 H \dot \phi + m^2 \phi = 0,
   \label{BG-EOM}
\eea where the angular bracket indicates averaging over time scale
of order $m^{-1}$, see Eq.\ (5) in \cite{Axion-1997}. Under an {\it
ansatz} \bea
   & & \phi (t) = \phi_{+} (t) \sin{(mt)}
       + \phi_{-} (t) \cos{(mt)},
   \label{ansatz-BG}
\eea ignoring ${H / m}$ higher order terms, Eq.\ (\ref{BG-EOM})
leads to an approximate solution \cite{Ratra-1991} \bea
   & & \phi (t) = a^{-3/2} \left[ \phi_{+0} \sin{(mt)}
       + \phi_{-0} \cos{(mt)} \right],
   \label{BG-phi}
\eea where $\phi_{+0}$ and $\phi_{-0}$ are constant coefficients.
Thus, \bea
   & & \mu = {1\over 2} m^2 a^{-3} \left( \phi_{+0}^2 + \phi_{-0}^2 \right), \quad
       p = 0,
\eea and the background medium evolves exactly same as a
pressureless ideal fluid \cite{Turner-1983}.

We consider only the scalar-type perturbations. Our conventions are
\cite{Bardeen-1988} \bea
   & & d s^2 = - \left( 1 + 2 \alpha \right) d t^2
       - 2 a \beta_{,\alpha} dt d x^\alpha
   \nonumber \\
   & & \quad
       + a^2 \left[ \left( 1 + 2 \varphi \right) \delta_{\alpha\beta}
       + 2 \gamma_{,\alpha\beta}
       \right] d x^\alpha d x^\beta,
   \\
   & & T^0_0 = - \mu - \delta \mu, \quad
       T^0_\alpha = - {1 \over k} \left( \mu + p \right)
       v_{,\alpha},
   \nonumber \\
   & & T^\alpha_\beta
       = \left( p + \delta p \right) \delta^\alpha_\beta
       + {1 \over a^2} \left( \nabla^\alpha \nabla_\beta
       - {1 \over 3} \Delta \delta^\alpha_\beta \right) \sigma.
\eea To the linear order, both the vector-type (rotation) and the
tensor-type (gravitational waves) perturbation equations are not
directly affected by the presence of the minimally coupled scalar
field including the axion \cite{Bardeen-1988}. The basic
perturbation equations we need are the Raychaudhury equation, the
energy-conservation equation, and the momentum conservation
equation, respectively \cite{Bardeen-1988} \bea
   & & \dot \kappa + 2 H \kappa
       + \left( 3 \dot H - {k^2 \over a^2} \right) \alpha
       = 4 \pi G \left( \delta \mu + 3 \delta p \right),
   \label{eq-4} \\
   & & \delta \dot \mu + 3 H \left( \delta \mu + \delta p \right)
       = \left( \mu + p \right) \left( \kappa - 3 H \alpha
       - {k \over a} v \right),
   \label{eq-6} \\
   & & {[a^4 (\mu + p) v]^\cdot \over a^4(\mu + p)}
       = {k \over a} \alpha
       + {k \over a (\mu + p)} \left( \delta p
       - {2 \over 3} {k^2 \over a^2} \sigma \right).
   \label{eq-7}
\eea The temporal average of oscillating scalar field contributes to
perturbed fluid quantities as \cite{Bardeen-1988} \bea
   & & \delta \mu = \langle \dot \phi \delta \dot \phi
       - \dot \phi^2 \alpha
       + m^2 \phi \delta \phi \rangle,
   \nonumber \\
   & & \delta p = \langle \dot \phi \delta \dot \phi
       - \dot \phi^2 \alpha
       - m^2 \phi \delta \phi \rangle,
   \nonumber \\
   & & {a \over k} \left( \mu + p \right) v
       = \langle \dot \phi \delta \phi \rangle, \quad
       \sigma = 0.
   \label{fluid-pert}
\eea For the scalar field we have the equation of motion
\cite{Bardeen-1988} \bea
   & & \delta \ddot \phi + 3 H \delta \dot \phi
       + {k^2 \over a^2} \delta \phi + V_{,\phi\phi} \delta \phi
   \nonumber \\
   & & \quad
       = \dot \phi \left( \kappa + \dot \alpha \right)
       + \left( 2 \ddot \phi + 3 H \dot \phi \right) \alpha.
   \label{eq-scalar}
\eea The above set of equations is spatially gauge-invariant. But we
have not taken the temporal gauge (hypersurface) condition which
will be used as an advantage in handling the mathematical analyses.

For $\delta \phi$ we take an {\it ansatz} \bea
   \delta \phi (k, t) = \delta \phi_{+} (k, t) \sin{(mt)}
       + \delta \phi_{-}  (k, t) \cos{(mt)}.
   \label{ansatz}
\eea Using Eqs.\ (\ref{BG-phi}) and (\ref{ansatz}), Eq.\
(\ref{fluid-pert}) gives to the leading order in $H/m$ \bea
   & & \delta \mu
       = a^{-3/2} m \Big[ m \left( \phi_{+0} \delta \phi_{+}
       + \phi_{-0} \delta \phi_{-} \right)
   \nonumber \\
   & & \quad
       + {1\over 2} \left( \phi_{+0} \delta \dot \phi_{-}
       - \phi_{-0} \delta \dot \phi_{+} \right) \Big] - \mu \alpha,
   \nonumber \\
   & & \delta p
       = {1\over 2} a^{-3/2} m \left( \phi_{+0} \delta \dot \phi_{-}
       - \phi_{-0} \delta \dot \phi_{+} \right) - \mu \alpha,
   \nonumber \\
   & & {a \over k} \left( \mu + p \right) v
       = {1\over 2} a^{-3/2} m \left( \phi_{+0} \delta \phi_{-}
       - \phi_{-0} \delta \phi_{+} \right).
   \label{fluid-2}
\eea We have not taken the temporal gauge condition yet.

%%%%%%%%%%%%%%%%%%%%%%%%%%%%%%%%%%%%%%%%%%%%%%%%%%%%%%%%%%%%%%%
The comoving gauge  takes $v = 0$ as the temporal gauge
(hypersurface) condition; in the presence of additional fluid, like
dust, radiation, neutrinos, etc., our gauge corresponds to the
axion-comoving gauge. Equation (\ref{eq-7}) gives \bea
   & & \alpha
       = - {\delta p \over \mu}.
   \label{alpha-CG}
\eea Equation (\ref{eq-6}), and Eqs.\ (\ref{eq-4}) and (\ref{eq-7}),
respectively, give \bea
   & & \dot \delta = \kappa,
   \label{kappa-eq-CG} \\
   & & \dot \kappa + 2 H \kappa
       = 4 \pi G \mu \delta
       - {k^2 \over a^2} {\delta p \over \mu},
   \label{dot-delta-eq-CG}
\eea where $\delta \equiv \delta \mu/\mu$. Combining these equations
we have \bea
   & & \ddot \delta + 2 H \dot \delta
       - 4 \pi G \mu \delta
       + {k^2 \over a^2} {\delta p \over \mu} = 0.
   \label{ddot-delta-eq}
\eea This equation is well known in the Newtonian context, see Eq.\
(15.9.23) in \cite{Weinberg-1972}, and also valid in Einstein
gravity context in the limit of negligible pressure
\cite{Lifshitz-1946}. In the case of CDM we effectively have $\delta
p = 0$. However, the equation of state relating $\delta p$ with
$\delta \mu$ in the case of axion is not determined at this point.
The equation of state of axion will follow from the perturbed
equation of motion.

Under the comoving gauge, Eq.\ (\ref{fluid-2}) gives \bea
   & & {\delta \phi_- \over \phi_{-0}}
       = {\delta \phi_+ \over \phi_{+0}}.
\eea Relation between $\delta p$ and $\delta \mu$ can be determined
through $\alpha$ using Eq.\ (\ref{eq-scalar}). We solve Eq.\
(\ref{eq-scalar}) strictly to leading order in ${H / m}$ as the
solution for the background is valid to such an order. Using Eqs.\
(\ref{alpha-CG}) and (\ref{kappa-eq-CG}), Eq.\ (\ref{eq-scalar})
gives \bea
   & & \alpha = - {1 \over 2} a^{3/2}
       {\delta \phi_+ \over \phi_{+0}} {k^2 \over m^2 a^2}.
\eea Equation (\ref{fluid-2}) gives \bea
   & & \delta = 2 a^{3/2} {\delta \phi_+ \over \phi_{+0}}
       \left( 1 + {1 \over 4} {k^2 \over m^2 a^2} \right).
\eea Therefore, we have an equation of state \bea
   & & \delta p = {1 \over 4} {k^2 \over m^2 a^2}
       {1 \over 1 + {1 \over 4} {k^2 \over m^2 a^2}}
       \delta \mu.
   \label{EOS}
\eea The effective sound speed becomes \bea
   & & c_s \equiv \sqrt{\delta p \over \delta \mu}
       = {1 \over 2} {k \over ma}
       \left( 1 + {1 \over 4} {k^2 \over m^2 a^2} \right)^{-1/2},
   \label{c_s}
\eea which shows an interesting scale dependence; for $k/(ma) \ll 1$
the time dependence $c_s \propto 1/a$ is the same as in the ordinary
matter dominated medium, see later. Equation (\ref{ddot-delta-eq})
becomes \bea
   \ddot \delta + 2 H \dot \delta
       - \left( 4 \pi G \mu
       - {1 \over 4} {k^4 \over m^2 a^4}
       {1 \over 1 + {1 \over 4} {k^2 \over m^2 a^2}}
       \right) \delta = 0.
   \label{ddot-delta-eq2}
\eea We note that we only have assumed $(H/m)^2 \ll 1$, but have not
assumed any condition on $(k/aH)^2$. Thus our equations are valid in
all scales. Even in the small scales where the pressure gradient
term has a role, in general we have $k^4/(m^2 a^4 H^2) \gg k^2/(m^2
a^2)$, thus ignoring $k^2/(m^2 a^2)$ term we have \bea
   \ddot \delta + 2 H \dot \delta
       - \left( 4 \pi G \mu
       - {1 \over 4} {k^4 \over m^2 a^4}
       \right) \delta = 0.
   \label{ddot-delta-eq3}
\eea This form was presented in Eq.\ (20) of
\cite{Sikivie-Yang-2009} based on the zero-shear gauge (often termed
the conformal Newtonian gauge or the longitudinal gauge). Although
Eq.\ (\ref{ddot-delta-eq3}), more precisely Eq.\
(\ref{ddot-delta-eq2}), in our comoving gauge is valid in all
scales, in the zero-shear gauge it is valid only for $k/(aH) \gg 1$;
the general density perturbation equation of pressureless medium in
the zero-shear gauge is quite complicated, see Eq.\ (27) in
\cite{HN-Newtonian-1999}. The competition between gravity and
pressure gradient terms in Eq.\ (\ref{ddot-delta-eq3}) gives the
Jeans scale; assuming a flat axion dominated model we have currently
\bea
   & & \lambda_J \equiv {2 \pi a \over k_J}
       \equiv 2 \pi \left( 16 \pi G \mu m^2 \right)^{-1/4}
   \nonumber \\
   & & \quad
       = 5.4 \times 10^{14} cm h^{-1/2}
       \left( {m \over 10^{-5} eV} \right)^{-1/2},
   \label{Jeans-scale}
\eea which is quite small corresponding to the solar-system size for
$m \sim 10^{-5} eV$, see also \cite{Sikivie-Yang-2009}. On scales
larger than $\lambda_J$, thus effectively in all cosmologically
relevant scales, the axion fluid can be regarded as the pressureless
ideal fluid. Therefore, the axion is justified as a CDM candidate.

Equation (\ref{ddot-delta-eq3}) can be analytically solved for
$\Lambda = 0$. For the background, we have $a \propto t^{2/3}$, $H =
2/(3 t)$, and $\mu = 1/(6 \pi G t^2)$. Equation
(\ref{ddot-delta-eq3}) has an exact solution \bea
   \delta_{\pm} \propto t^{-1/6} J_{\mp 5/2}
       ( 3 A t^{-1/3} ), \quad
       A \equiv {k^2 \over 2m} \left( {t^{2/3} \over a}
       \right)^2.
   \label{exact-solution}
\eea This can be compared with an exact solution in matter dominated
era with an equation of state $p \propto \mu^\gamma$
\cite{Weinberg-1972}. For $\gamma = {5 / 3}$ we have the solution in
Eq.\ (\ref{exact-solution}) with $A = c_s k t^{4/3} / a$, see Eq.\
(15.9.41) in \cite{Weinberg-1972}; notice that $c_s \propto 1/a$
even in the matter dominated medium. Thus, again we can identify
$c_s = k/(2ma)$ as an effective sound speed of the axion fluid.

For ${k^2 / (m a H)} \ll 1$ and $\Lambda = 0$ we have a perturbative
solution \bea
   & & \delta (k, t)
       = c_+ (k) t^{2/3} \left[
       1 + {3 \over 8} {k^4 \over m^2}
       \left( {t^{2/3} \over a} \right)^4 t^{-2/3} \right]
   \nonumber \\
   & & \quad
       + c_- (k) t^{-1} \left[
       1 - {9 \over 56} {k^4 \over m^2}
       \left( {t^{2/3} \over a} \right)^4 t^{-2/3} \right].
\eea This coincides with the solution in Eq.\ (25) of
\cite{Axion-1997}. Our previous proof of the axion as a CDM
candidate presented in \cite{Axion-1997} was made based on this and
other solutions.

%%%%%%%%%%%%%%%%%%%%%%%%%%%%%%%%%%%%%%%%%%%%%%%%%%%%%%%%%%%%%%%%%%%%
%\section{Discussion}

For a clear presentation, in this work we have considered a flat
background composed of a single axion component but including the
cosmological constant. Our analysis can be easily extended to
non-flat case as well as in the presence of additional fluids and
fields. We can show that axion behaves as a CDM even in such more
realistic situations. For example, in the presence of other
components the equation of state in Eq.\ (\ref{EOS}) is valid for
the axion component.

Equation (\ref{Jeans-scale}) shows an axion Jeans scale where linear
perturbation of axion fluid becomes stabilized (oscillates) under
that scale. The axion fluid shows peculiar equation of state and
effective sound speed presented in Eqs.\ (\ref{EOS}) and
(\ref{c_s}). Although our neighborhood in the solar system is in a
significantly nonlinear stage it is curious to see the possible
observational signature of axion fluid based on its contribution to
effective pressure. This is left for future investigation.
%\textbf{Professor Sikivie: do you have any suggestion on this point?}

Recently we have shown that even to the second-order perturbation,
the density and velocity perturbation equations of the zero-pressure
medium in Einstein's gravity exactly coincide with the ones in
Newton's gravity: we call this a relativistic/Newtonian
correspondence \cite{second-order}. The relativistic/Newtonian
correspondence to the linear order can be found in Eq.\
(\ref{ddot-delta-eq}) with vanishing pressure. In the present work
we have shown that the axion properly treated in relativistic
perturbation theory behaves as a pressureless fluid in
cosmologically relevant scales, thus justified as a CDM candidate:
we may call it the axion/CDM correspondence to the linear order.
Whether such a correspondence continues even in the case of
second-order perturbation is an interesting open issue at the
moment. We will address this important issue in a future occasion.

\vskip .0cm
%%%%%%%%%%%%%%%%%%%%%%%%%%%%%%%%%%%%%%%%%%%%%%%%%%%%%%%%%%%%%%%
%
% Acknowledgments
%
%%%%%%%%%%%%%%%%%%%%%%%%%%%%%%%%%%%%%%%%%%%%%%%%%%%%%%%%%%%%%%%
\noindent{\bf Acknowledgments:} J.H.\ wishes to thank Professor
Pierre Sikivie for useful discussion in Aspen winter conference
which initiated this work. H.N.\ was supported by grant No. C00022
from the Korea Research Foundation. J.H.\ was supported by the Korea
Research Foundation Grant funded by the Korean Government (MOEHRD,
Basic Research Promotion Fund) (No.\ KRF-2007-313-C00322)
(KRF-2008-341-C00022), and by Grant No.\ R17-2008-001-01001-0 from
the Korea Science and Engineering Foundation.

%%%%%%%%%%%%%%%%%%%%%%%%%%%%%%%%%%%%%%%%%%%%%%%%%%%%%%%%%%%%%%%
%
% Bibliography
%
%%%%%%%%%%%%%%%%%%%%%%%%%%%%%%%%%%%%%%%%%%%%%%%%%%%%%%%%%%%%%%%

%%%%%%%%%%%%%%%%%%%%%%%%%%%%%%%%%%%%%%%%%%%%%%%%%%%%%%%%%%%%%%
\end{document}